\documentclass[11pt]{article}

\usepackage{amsmath}
\usepackage{graphicx}
\usepackage{indentfirst}
\usepackage{amssymb}
\usepackage{cite}
\usepackage{color}
\usepackage{subfigure}
\usepackage{varwidth}
\usepackage[colorlinks=true, linkcolor=red, citecolor=blue, urlcolor=magenta]{hyperref}
\usepackage{xcolor}

\begin{document}
\title{Polarization Images of Neutron Stars Illuminated by a Thin Accretion Disk: A Comparison with Black Holes}

\date{}
\maketitle
\begin{center}
\author{Dan-Dan Feng,}$^{a}$\footnote{E-mail: ncddfeng@163.com}
\author{Guo-Ping Li,}$^{b}$\footnote{E-mail: gpliphys@yeah.net(Corresponding author)}
\author{Chen-Yu Yang,}$^{c}$\footnote{E-mail: chenyuyang\_2024@163.com(Corresponding author)}
\author{Hong Lei}$^{d}$\footnote{E-mail: 405114039@qq.com}
\\

\vskip 0.25in
$^{a}$\it{School of Mathematics, Physics and Statistics, Sichuan Minzu College, \\ Ganzi 626100, China}\\
$^{b}$\it{Physics and Astronomy College, China West Normal University,\\ Nanchong 637000, China}\\
$^{c}$\it{Department of Mechanics, Chongqing Jiaotong University, Chongqing 400000,  China}\\
$^{d}$\it{Chongqing Bashu Science City  Secondary School,Chongqing 401331, China}\\
\end{center}
\vskip 0.6in
{\abstract
{
We investigate linear polarization images of static, spherically symmetric neutron stars illuminated by a geometrically and optically thin accretion disk. Neutron star equilibrium configurations are constructed with a polytropic equation of state, and the null geodesic equations and the parallel transport equation for the linear polarization vector are solved in the geometric optics approximation. The numerical results show that the total polarized intensity is generally positively correlated with the total intensity and reaches its maximum near the neutron star surface. As the observer inclination increases, the symmetry of the polarization images is progressively broken. In addition, the magnetic field configuration mainly affects the direction of the polarization vectors, while its influence on the overall polarized intensity distribution is comparatively limited. To further characterize the spatial structure of the polarization direction, we introduce the net electric vector position angle $\chi_{\mathrm{net}}$ and the second azimuthal Fourier mode $\angle\beta_2$. A comparison with polarization images of a Schwarzschild black hole reveals clear differences between the two types of compact objects in the locations of strongly polarized regions and the size of the central region without a polarization signal. These results show that linear polarization images provide information beyond total intensity images for distinguishing neutron stars from black holes.
}}

\thispagestyle{empty}
\newpage
\setcounter{page}{1}

\section{Introduction}

In recent years, the Event Horizon Telescope (EHT) Collaboration has released the $230~\mathrm{GHz}$ shadow images of M87* and of Sgr~A* at the Galactic center~\cite{event2019first,akiyama2022first}. It subsequently published their polarization images~\cite{akiyama2021first,akiyama2024first}. These observations revealed ring-like emission structures on scales comparable to the event horizon. The images are consistent with theoretical expectations for Kerr black holes within general relativity (GR), while the polarization images provide further information about the magnetic field and emission structure near the black holes. Imaging on event horizon scales and polarimetric observations have therefore become important tools for studying strong gravity, accretion flows, and the immediate environments of compact objects. Black holes and neutron stars are important astrophysical laboratories for testing gravity in the strong-field regime~\cite{berti2015testing}. Exotic compact objects such as boson stars provide possible alternative models for testing the black hole paradigm~\cite{cardoso2019testing,eichhorn2023horizonless,wang2023different}. These studies have advanced compact object astrophysics and established an observational basis for comparing the optical and polarization signatures of different compact objects.

Among compact objects, neutron stars are distinguished from black holes by their material surfaces. These compact stars form during the late stages of massive stellar evolution. A typical neutron star has a mass of approximately $1.5M_{\odot}$, a radius of approximately $12\,\mathrm{km}$, and a central density several times the nuclear saturation density~\cite{lattimer2004physics}. In early work on neutron stars, Baade and Zwicky proposed that a supernova might mark the transition of an ordinary star into a neutron star~\cite{baade1934cosmic}. The discovery of a rapidly pulsating radio source~\cite{hewish1968observation} and the proposal of the rotating neutron star model~\cite{gold1968rotating} established the observational and theoretical foundations of pulsar studies. The macroscopic mass--radius relation of a neutron star is determined by its internal equation of state (EOS). Measurements of the mass and radius can therefore constrain the composition and properties of matter at supranuclear densities~\cite{ozel2016masses}. Measurements of the Shapiro delay in a binary system confirmed a neutron star with a mass of $(1.97\pm0.04)M_{\odot}$ and ruled out most EOSs then available that contained hyperons or boson condensates~\cite{demorest2010two}. An energy-dependent thermal X-ray waveform analysis of PSR~J0030+0451 with NICER provided estimates of the pulsar mass and radius~\cite{miller2019psr}. Beyond measurements of mass and radius, high-resolution optical images of neutron stars may also contain information about their surface scale, exterior spacetime, and surrounding radiative environment.

The shadows and optical images of accreting compact objects have been studied extensively~\cite{zeng2022qed,gao2023investigating,zeng2023optical}. As early as 1979, Luminet used a semianalytic method to simulate the image of a Schwarzschild black hole illuminated by a thin accretion disk~\cite{luminet1979image}. Gralla et al. later introduced a geometrically and optically thin accretion disk model. In this emission model, a Schwarzschild black hole image can be decomposed into a direct image, a lensing ring, and a photon ring~\cite{gralla2019black}. The model has subsequently been applied to investigate the effects of dark energy, quantum corrections, torsion charge, and global monopoles on images of spherically symmetric black holes~\cite{zeng2020influence,peng2021influence,he2022influence,li2021observational}. Hou et al. extended thin disk imaging to a rotating Kerr--Melvin black hole in an external magnetic field and explored whether the inner shadow and critical curve could be used to estimate the magnetic field strength~\cite{hou2022image}. Related studies have also considered images of regular black holes, black holes with effective corrections from quantum gravity, and rotating black holes in dark matter environments~\cite{zeng2023holographic,he2025observational,yang2025shadow,li2025shadow,zeng2025kerr}. Beyond black holes, horizonless compact objects such as boson stars and Proca stars illuminated by thin accretion disks may exhibit multiple bright rings, central dark regions, or central emission regions~\cite{yang2026observational2,li2026observational,rosa2022shadows,he2025observation}. In our previous work, we analyzed optical images of neutron stars illuminated by thin and thick accretion disks~\cite{yang2026distinguishing,yang2026distinguishing2}. For a thin disk, the intensity reaches its maximum at the stellar surface. For a thick disk, the higher-order images of a neutron star are larger than those of a black hole with the same parameter settings. These studies demonstrate that both the type of compact object and the accretion disk model affect the optical image.

Most existing studies focus on shadows and emission intensity distributions, whereas polarization images can provide additional information about the magnetic field geometry and matter motion in the emitting region. The EHT polarization images of M87* and Sgr~A* reveal pronounced polarization patterns within the emission rings. The electric vector position angle (EVPA) pattern of M87* is approximately azimuthal~\cite{akiyama2021first}, while Sgr~A* exhibits a pronounced spiral polarization structure~\cite{akiyama2024first}. Constructing polarization images generally requires tracing radiation along null geodesics and calculating the parallel transport of the polarization vector~\cite{aimar2024gyoto,yang2026shadow}. In simplified models, this procedure can be implemented with analytic approximations. Narayan et al. modeled an axisymmetric equatorial emitting ring around a Schwarzschild black hole. They combined Beloborodov's approximate analytic expression for light propagation~\cite{beloborodov2002gravitational} with the Walker--Penrose constant to obtain analytic estimates of the image polarization. For suitable parameters, this model can reproduce the EVPA pattern and relative polarization intensity in images of M87*~\cite{narayan2021polarized}. Gelles et al. constructed a simplified model with an equatorial emission source and generated the corresponding polarization images of a Kerr black hole~\cite{gelles2021polarized}. These results show that the polarization signatures are jointly determined by the magnetic field geometry, the intrinsic black hole parameters, and the observer inclination. Polarization images of other black holes, naked singularities, and horizonless compact objects have also been studied extensively~\cite{qin2022polarized,deliyski2023polarized,yin2026bright,zeng2026polarization}, providing an effective approach to probing accretion physics and the underlying spacetime geometry. However, systematic studies of neutron star polarization images remain limited. Compared with emission intensity images alone, polarization images can further reveal how the magnetic field configuration affects the observable signatures and thereby provide a more complete description of the physical environment around a neutron star.

Motivated by the importance of polarization images, we consider neutron stars described by a polytropic EOS~\cite{flanagan2008constraining,lattimer2007neutron,hinderer2008tidal,read2009constraints}. We systematically investigate their polarization structure when illuminated by a geometrically and optically thin accretion disk and compare their polarization images with those of black holes to identify potential distinguishing features. The remainder of this paper is organized as follows. Sec.~\ref{sec2} introduces the equilibrium structure of neutron stars and the ray-tracing framework. Sec.~\ref{sec3} presents the construction and parallel transport of the polarization vector and defines the relevant polarization observables. Sec.~\ref{sec4} presents and analyzes the numerical results. Sec.~\ref{sec5} summarizes and discusses our findings. We use units in which $c=G=\hbar=1$, where $c$ is the speed of light in vacuum, $G$ is the gravitational constant, and $\hbar$ is the reduced Planck constant.

\section{Basic Setup}\label{sec2}

\subsection{Equilibrium Structure}

The spacetime metric of a static, spherically symmetric star takes the form~\cite{oppenheimer1939massive}
\begin{equation}
	ds^{2}
	=-e^{\nu(r)}dt^{2}
	+e^{\mu(r)}dr^{2}
	+r^{2}d\Omega^{2},
	\qquad
	d\Omega^{2}=d\theta^{2}+\sin^{2}\theta\,d\phi^{2},\label{ds2}
\end{equation}
where $\nu$ and $\mu$ depend only on the radial coordinate $r$. The stellar matter is modeled as a perfect fluid with the energy-momentum tensor
\begin{equation}
	T_{\alpha\beta}
	=(P+\epsilon)u_{\alpha}u_{\beta}+Pg_{\alpha\beta},
	\qquad
	u^{\alpha}
	=\left(e^{-\nu(r)/2},0,0,0\right),
\end{equation}
where $u^{\alpha}$ is the four-velocity and satisfies the normalization condition $u^{\alpha}u_{\alpha}=-1$. The quantities $P(r)$ and $\epsilon(r)$ denote the fluid pressure and energy density, respectively. Introducing the mass function $m(r)$, the metric function can be expressed as
\begin{equation}
	e^{-\mu(r)}=1-\frac{2m(r)}{r}, \quad
	m(r)=4\pi\int_{0}^{r}\epsilon(x)x^{2}\,dx.
\end{equation}
The $tt$ and $rr$ components of the Einstein equations give
\begin{align}
	\frac{dm}{dr}
	&=4\pi r^{2}\epsilon,\\
	\frac{d\nu}{dr}
	&=\frac{2\left(m+4\pi r^{3}P\right)}
	{r(r-2m)}.
\end{align}
The pressure also obeys the Tolman--Oppenheimer--Volkoff (TOV) equation
\begin{equation}
	\frac{dP}{dr}
	=-\frac{(P+\epsilon)\left(m+4\pi r^{3}P\right)}
	{r(r-2m)}.
\end{equation}
To perform the numerical integration, an EOS must be specified. For compact objects such as neutron stars, the EOS can be expressed as $f(P,\epsilon)=0$~\cite{liang2023differential}. We adopt the polytropic equation of state~\cite{read2009constraints,hinderer2008tidal},
\begin{equation}
	P-K\epsilon^{\Gamma}=0,
	\qquad
	\Gamma=1+\frac{1}{n_{\mathrm{p}}},
\end{equation}
where $K$ is a constant and $n_{\mathrm{p}}$ is the polytropic index. Numerically solving these equations yields the interior metric. The boundary condition $P(R_{\star})=0$ determines the stellar radius $R_{\star}$, and the corresponding mass is $M_{\star}=m(R_{\star})$. The exterior spacetime is described by the Schwarzschild solution. At the stellar surface, it obeys the matching condition
\begin{equation}
	e^{\nu(R_{\star})}
	=1-\frac{2M_{\star}}{R_{\star}}.
\end{equation}
The metric obtained by numerical integration is not used directly in the subsequent imaging calculation. Instead, we use a fitted representation of the metric. Further details can be found in Refs.~\cite{rosa2022shadows,yang2026distinguishing,he2025observation}.

\subsection{Ray-Tracing Framework}

We now briefly outline the ray-tracing method used in the imaging calculation. We define
\begin{equation}
	A(r)=e^{\nu(r)},\qquad B(r)=e^{-\mu(r)}.
\end{equation}
The metric in Eq.~\eqref{ds2} can then be written as
\begin{equation}
	ds^{2}
	=-A(r)dt^{2}
	+\frac{dr^{2}}{B(r)}
	+r^{2}\left(d\theta^{2}+\sin^{2}\theta\,d\phi^{2}\right).
\end{equation}
Let $x^{\alpha}=(t,r,\theta,\phi)$ and let $\lambda$ denote the affine parameter, with an overdot denoting differentiation with respect to $\lambda$. The Lagrangian for a null geodesic is
\begin{equation}
	L_{\mathrm{geo}}
	=\frac{1}{2}g_{\alpha\beta}\dot{x}^{\alpha}\dot{x}^{\beta}
	=\frac{1}{2}\left[
	-A(r)\dot{t}^{2}
	+\frac{\dot{r}^{2}}{B(r)}
	+r^{2}\dot{\theta}^{2}
	+r^{2}\sin^{2}\theta\,\dot{\phi}^{2}
	\right]
	=0,
\end{equation}
The equations of motion are
\begin{equation}
	\frac{d}{d\lambda}
	\left(\frac{\partial L_{\mathrm{geo}}}{\partial\dot{x}^{\alpha}}\right)
	-\frac{\partial L_{\mathrm{geo}}}{\partial x^{\alpha}}
	=0.
\end{equation}
Because the spacetime is spherically symmetric, the orbital plane of a null geodesic can be chosen as the equatorial plane without loss of generality, corresponding to $\theta=\pi/2$. Since the metric is independent of $t$ and $\phi$, the conserved energy and angular momentum of the photon are, respectively,
\begin{equation}
	E=-p_{t}=A(r)\dot{t},
	\qquad
	L_{\phi}=p_{\phi}=r^{2}\dot{\phi}.
\end{equation}
We define the impact parameter $b=L_{\phi}/E$, choose positive angular momentum, and set $L_{\phi}=1$. The tangent vector to the null geodesic is
\begin{align}
	\dot{x}^{\alpha}
	&=\left(\dot{t},\dot{r},\dot{\theta},\dot{\phi}\right) \nonumber\\
	&=\left(\frac{1}{A(r)b},\sqrt{B(r)\left[\frac{1}{A(r)b^{2}}-\frac{1}{r^{2}}\right]},0,\frac{1}{r^{2}}\right).
\end{align}

To obtain polarization images of the neutron star, we must choose an observer and construct an image plane. A natural choice is a zero angular momentum observer (ZAMO) located at $(t_{\mathrm{o}},r_{\mathrm{o}},\theta_{\mathrm{o}},\phi_{\mathrm{o}})$. In the present spacetime, the local orthonormal tetrad can be chosen as~\cite{velasquez2022osiris,grenzebach2014photon},
\begin{align}
	e_{(t)} & = \left(\frac{1}{\sqrt{A(r_{\mathrm{o}})}},0,0,0\right), \quad
	e_{(r)} = \left(0,-\sqrt{B(r_{\mathrm{o}})},0,0\right), \\
	e_{(\theta)} & = \left(0,0,\frac{1}{r_{\mathrm{o}}},0\right), \quad
	e_{(\phi)} = \left(0,0,0,-\frac{1}{r_{\mathrm{o}}\sin\theta_{\mathrm{o}}}\right).
\end{align}
In the ZAMO frame, the photon four-momentum is projected onto the local orthonormal tetrad as
\begin{equation}
	p_{(\alpha)} = p_\beta e^{\beta}_{(\alpha)},
\end{equation}
where $p_{(\alpha)}$ and $p_\beta$ denote the four-momentum components in the local ZAMO tetrad and the coordinate basis, respectively. At the observer's position, we introduce the celestial coordinates $(X,Y)$,
\begin{equation}
	\cos X = \frac{p^{(r)}}{p^{(t)}}, \qquad
	\tan Y = \frac{p^{(\phi)}}{p^{(\theta)}}.
\end{equation}
The Cartesian coordinates $(\tilde{\alpha},\tilde{\beta})$ on the image plane are obtained from the celestial coordinates by stereographic projection~\cite{hu2021qed},
\begin{equation}
	\tilde{\alpha} = -2 \tan\frac{X}{2}\,\sin Y, \qquad
	\tilde{\beta} = -2 \tan\frac{X}{2}\,\cos Y.
\end{equation}

In the ray-tracing calculation, we use the neutron star radius $R_{\star}$ as the inner termination boundary. Integration stops when a ray reaches $r=R_{\star}$, and we accumulate no contribution to either intensity or linear polarization from the stellar interior. This treatment is based on three considerations. First, the neutron star interior consists of high-density nuclear matter and is optically opaque to the electromagnetic radiation considered here. Photons entering the star undergo absorption and multiple scattering, which make it difficult to preserve their original polarization information. Second, we study photon propagation only in the exterior spacetime and focus on the effects of spacetime curvature on the geodesics and parallel transport of linear polarization. We do not construct a complete radiative transfer model for the stellar interior or surface layers. A realistic neutron star atmosphere or condensed surface may produce thermal radiation and polarized signals~\cite{ozel2013surface,potekhin2014atmospheres,pavlov2000polarization}, but such surface emission processes lie beyond the scope of this work. Third, terminating the geodesic and polarization transport at $R_{\star}$ prevents an inappropriate extension of the exterior propagation model into the stellar interior. This avoids unphysical integration outcomes and improves the numerical stability of the imaging and polarization calculations. Thus, $R_{\star}$ acts as an absorbing, optically thick boundary in our model.

\section{Linear Polarization Model and Observables}\label{sec3}

Linear polarization images provide more information about the radiation and geometry than images of intensity alone. In this section, we adopt the geometrically and optically thin accretion disk model introduced by Gralla et al.~\cite{gralla2019black} and investigate linearly polarized synchrotron radiation. In this model, the plasma in the thin accretion disk moves on timelike circular orbits, and its four-velocity satisfies the normalization condition
\begin{equation}
	g_{\mu\nu}u^{\mu}u^{\nu}=-1.
\end{equation}
For the static, spherically symmetric metric considered here, the four-velocity of the accretion flow takes the form~\cite{li2026observational}
\begin{equation}
	u_{\mu} =
	\left(
	-\sqrt{\frac{2g_{tt}^{2}}{r\partial_r g_{tt}-2g_{tt}}},
	0,
	0,
	\sqrt{\frac{r^{3}\partial_r g_{tt}}{2g_{tt}-r\partial_r g_{tt}}}
	\right).
\end{equation}
A light ray may intersect the thin accretion disk several times before reaching the observer. The total observed intensity must therefore include the contributions from all possible disk crossings. When reflection and absorption are neglected, it can be written as
\begin{equation}
	I_{\mathrm{obs}}=\sum_{n=1}^{N}g_n^{3}I_n , \label{eq:Io}
\end{equation}
where $I_n$ is the emissivity and $N$ is the maximum number of disk crossings. The redshift factor $g_n$ is defined as the ratio of the observed frequency to the emitted frequency and takes the form
\begin{equation}
	g_n=-\frac{1}{u_{\mu}p^{\mu}},
\end{equation}
where $p^{\mu}$ is the photon four-momentum. For synchrotron radiation, we adopt the emission profile $I_n$~\cite{gralla2020shape},
\begin{equation}
	I_n(r)=
	\frac{
		\exp\left[
		-\frac{1}{2}
		\left(
		c_1+\operatorname{arcsinh}\left(\frac{r-c_2}{c_3}\right)
		\right)^2
		\right]
	}{
		\sqrt{(r-c_2)^2+c_3^2}
	}.
\end{equation}
The radial profile of $I_n$ is controlled by the three parameters $(c_1,c_2,c_3)$. The parameter $c_1$ regulates the growth of the intensity, $c_2$ determines the location of its maximum, and $c_3$ controls the width of the profile. We set $c_1=0$, $c_2=6M_{\star}$, and $c_3=M_{\star}$ in the calculations below.

In the thin disk model, the plasma in the equatorial plane follows circular timelike geodesics. The linearly polarized light arises from synchrotron radiation emitted by electrons in the plasma. For an observer comoving with the plasma, the polarization direction of the emitted light is perpendicular to both the local magnetic field $\vec{B}$ and the spatial momentum $\vec{p}$ of the photon. The spatial components of the photon polarization vector can therefore be chosen as~\cite{hou2025near},
\begin{equation}
	\vec{f} = \frac{\vec{p} \times \vec{B}}{|\vec{p}|}.
\end{equation}
The corresponding generally covariant form of the photon polarization vector is
\begin{equation}
	f^\mu \propto \xi^{\mu\nu\lambda\eta} u_\nu p_\lambda B_\eta ,
\end{equation}
where $\xi^{\mu\nu\lambda\eta}$ is the Levi-Civita tensor. The quantities $u_\nu$, $p_\lambda$, and $B_\eta$ are the covariant components of the plasma four-velocity, photon four-momentum, and magnetic field four-vector, respectively. Once the direction of the polarization vector has been determined, it can be normalized to satisfy
\begin{equation}
	f^\mu f_\mu = 1.
\end{equation}
The intensities of linearly polarized light and natural light at the emission point are represented by the emission functions $I_{\mathrm{p}}$ and $I_{\mathrm{i}}$, respectively. For simplicity, we assume that the emission intensity is independent of the photon frequency and magnetic field and depends only on position. We further assume that the emitted light is completely linearly polarized, so that
\begin{equation}
	I_{\mathrm{i}}=I_{\mathrm{i}}(r),\qquad I_{\mathrm{p}}= I_{\mathrm{i}}(r).
\end{equation}
In the geometric optics approximation, the observed polarization properties are determined by parallel transport of the linear polarization vector $f^\mu$ along the null geodesic to the observer. The transport equation is
\begin{equation}
	p^\nu \nabla_\nu f^\mu=0,
\end{equation}
or equivalently,
\begin{equation}
	\frac{d}{d\lambda} f^\mu
	+\Gamma^\mu{}_{\nu\rho}p^\nu f^\rho=0.
\end{equation}
As in the unpolarized case, the observed linearly polarized intensity $P_{\nu_{\mathrm{o}}}$ and total intensity $I_{\nu_{\mathrm{o}}}$ can be written as
\begin{equation}
	P_{\nu_{\mathrm{o}}}=g^{3}I_{\mathrm{p}},
	\quad
	I_{\nu_{\mathrm{o}}}=g^{3}I_{\mathrm{i}}.
\end{equation}

We focus on two observables in the polarization images. The first is the total polarized intensity $P_{\mathrm{obs}}$, and the second is the direction of linear polarization, described by the electric vector position angle (EVPA) $\chi_{\mathrm{evpa}}$. The projected components of the polarization vector on the image plane satisfy
\begin{equation}
	f^{(\tilde{\alpha})}=f^\mu \cdot e_{\tilde{\alpha}}=-f^\mu \cdot e_\phi,\quad
	f^{(\tilde{\beta})}=f^\mu \cdot e_{\tilde{\beta}}=-f^\mu \cdot e_\theta.
\end{equation}
Because $\vec{f}$ and $-\vec{f}$ describe the same state of linear polarization, the EVPA is defined only modulo $\pi$. We choose $f^{(\tilde{\beta})}>0$ and $\chi_{\mathrm{evpa}}\in[0,\pi)$. The total polarized intensity observed at any point $(\tilde{\alpha},\tilde{\beta})$ on the image plane is determined by the linearly polarized radiation from all intersections of the ray with the equatorial plane. According to the definitions of the Stokes parameters $Q$ and $U$~\cite{huang2024coport}, the corresponding contributions to $Q$ and $U$ add linearly. We therefore have
\begin{align}
	Q_{\mathrm{all}}
	&=\sum_{n=1}^{N}Q_n
	=\sum_{n=1}^{N}g_n^3 I_{\mathrm{p}_n}
	\left[\left(f_n^{(\tilde{\alpha})}\right)^2-\left(f_n^{(\tilde{\beta})}\right)^2\right],\\
	U_{\mathrm{all}}
	&=\sum_{n=1}^{N}U_n
	=\sum_{n=1}^{N}g_n^3 I_{\mathrm{p}_n}
	\left(2f_n^{(\tilde{\alpha})}f_n^{(\tilde{\beta})}\right).
\end{align}
The observed total polarized intensity and EVPA are then, respectively,
\begin{equation}
	P_{\mathrm{obs}}=\sqrt{Q_{\mathrm{all}}^{2}+U_{\mathrm{all}}^{2}}, \quad
	\chi_{\mathrm{evpa}}
	=\frac{1}{2}\operatorname{atan2}
	\left(U_{\mathrm{all}},Q_{\mathrm{all}}\right).
\end{equation}

To further quantify the structure of the linear polarization direction on the image plane, we introduce two quantities based on the EVPA. These are the ``net EVPA'' $\chi_{\mathrm{net}}$~\cite{wang2026imaging,chen2025polarization} and the second azimuthal Fourier mode $\angle\beta_2$~\cite{palumbo2020discriminating,ricarte2022observational,wong2026black}. We use polar coordinates $(\rho,\varphi)$ on the image plane,
\begin{equation}
	\tilde{\alpha}=\rho\cos\varphi,\quad
	\tilde{\beta}=\rho\sin\varphi,
\end{equation}
where $\rho$ and $\varphi$ are the polar radius and polar angle, respectively. The ``net EVPA'' $\chi_{\mathrm{net}}$ is defined as
\begin{equation}
	\chi_{\mathrm{net}}=\chi_{\mathrm{evpa}}-\varphi.
\end{equation}
This quantity measures the deflection of the polarization vector from the radial direction on the screen. Because the direction of linear polarization has a $180^\circ$ degeneracy, we take $\chi_{\mathrm{net}}\in[0^\circ,180^\circ)$. With this convention, $\chi_{\mathrm{net}}=0^\circ$ or, equivalently, $180^\circ$ indicates a radial alignment of the polarization vectors, whereas $\chi_{\mathrm{net}}=90^\circ$ indicates an azimuthal alignment. The range $0^\circ<\chi_{\mathrm{net}}<90^\circ$ corresponds to a counterclockwise rotation of the polarization vectors from the radial direction. Conversely, $90^\circ<\chi_{\mathrm{net}}<180^\circ$ corresponds to a clockwise pattern.

The second azimuthal Fourier mode $\angle\beta_2$ characterizes the mean rotation of the polarization direction relative to the radial reference direction within an annular region. We divide the image plane into concentric annuli. Because only the phase of $\beta_2$ is relevant here, we define it as
\begin{equation}
	\angle\beta_2(\rho_k)
	=\arg\left[
	\sum_{(\tilde{\alpha},\tilde{\beta})\in\mathcal{A}_k}
	P_{\mathrm{obs}}(\tilde{\alpha},\tilde{\beta})
	\exp\left(2i\chi_{\mathrm{net}}(\tilde{\alpha},\tilde{\beta})\right)
	\right],
\end{equation}
Here, $\mathcal{A}_k$ denotes the $k$th annulus and $\rho_k$ is its representative radius. We take $\angle\beta_2$ to lie in $[-180^\circ,180^\circ]$. Within this range, $0^\circ<\angle\beta_2<180^\circ$ ($-180^\circ<\angle\beta_2<0^\circ$) indicates a mean counterclockwise (clockwise) rotation of the polarization direction relative to the radial direction. The values $\angle\beta_2=0^\circ$ and $\angle\beta_2=\pm180^\circ$ correspond to predominantly radial and azimuthal patterns, respectively. Unlike $\chi_{\mathrm{net}}$, $\angle\beta_2$ describes the mean rotation of the polarization direction within an annulus and is therefore better suited for comparing the overall polarization structure of neutron stars with different parameters.

\section{Polarization Images and Numerical Results}\label{sec4}

The polarization signatures of a neutron star are closely related to its magnetic field configuration. The magnetic field distribution needs to be specified first in the numerical calculation~\cite{hu2022polarized}. We parameterize the magnetic field in the comoving frame of the fluid as $\vec{B}=\vec{B}(B_r,B_\phi,B_z)$. The configurations $\vec{B}(B_r,0,0)$ and $\vec{B}(0,B_\phi,0)$ denote purely radial and purely azimuthal magnetic fields in the equatorial plane, respectively, while $\vec{B}(0,0,B_z)$ denotes a magnetic field perpendicular to that plane. Polarimetric simulations of M87* show that the configurations $\vec{B}(0.87,0.5,0)$ and $\vec{B}(0.97,0.26,0)$ can reproduce its polarization signatures in particular states~\cite{narayan2021polarized}. We therefore use $\vec{B}(0.87,0.5,0)$ as our primary configuration and briefly discuss several other magnetic field configurations.

\begin{figure}[!htbp]
	\centering 
	\subfigure[$\theta_{\mathrm{o}}=0^\circ,n_{\mathrm{p}}=1.1$]{\includegraphics[scale=0.4]{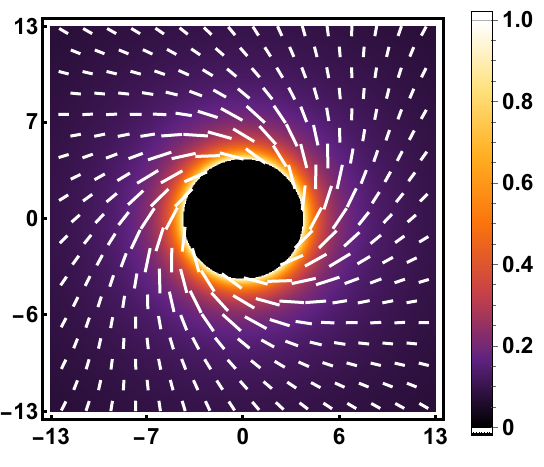}}
	\subfigure[$\theta_{\mathrm{o}}=20^\circ,n_{\mathrm{p}}=1.1$]{\includegraphics[scale=0.4]{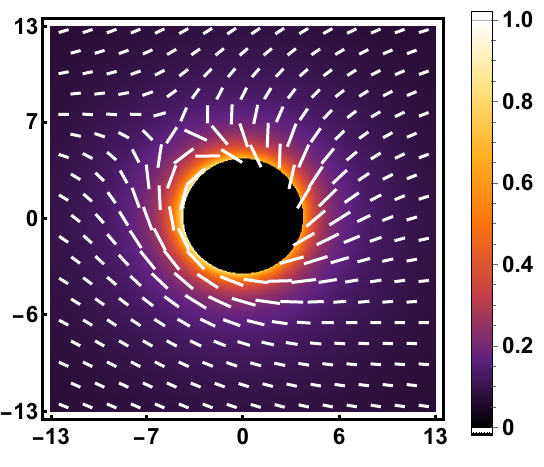}}
	\subfigure[$\theta_{\mathrm{o}}=55^\circ,n_{\mathrm{p}}=1.1$]{\includegraphics[scale=0.4]{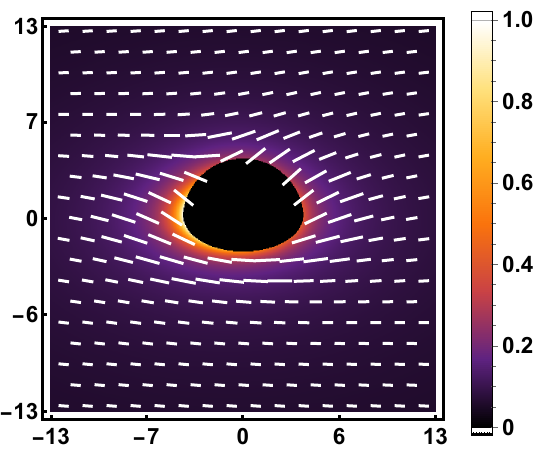}}
	\subfigure[$\theta_{\mathrm{o}}=80^\circ,n_{\mathrm{p}}=1.1$]{\includegraphics[scale=0.4]{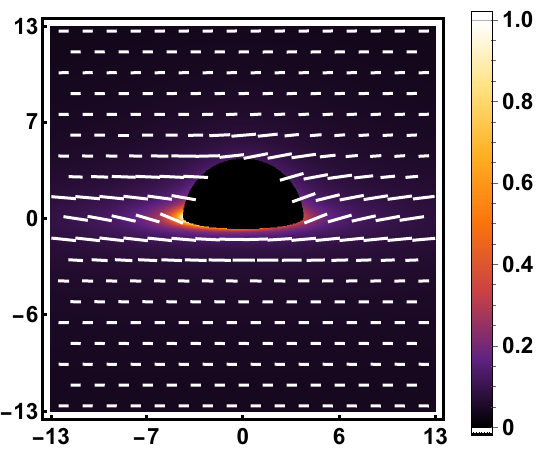}}
	
	\subfigure[$\theta_{\mathrm{o}}=0^\circ,n_{\mathrm{p}}=1.2$]{\includegraphics[scale=0.4]{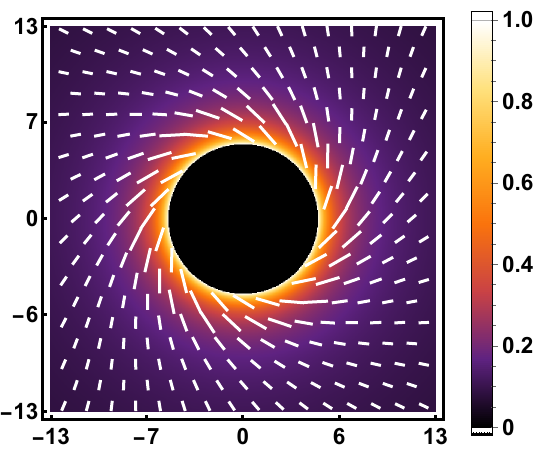}}
	\subfigure[$\theta_{\mathrm{o}}=20^\circ,n_{\mathrm{p}}=1.2$]{\includegraphics[scale=0.4]{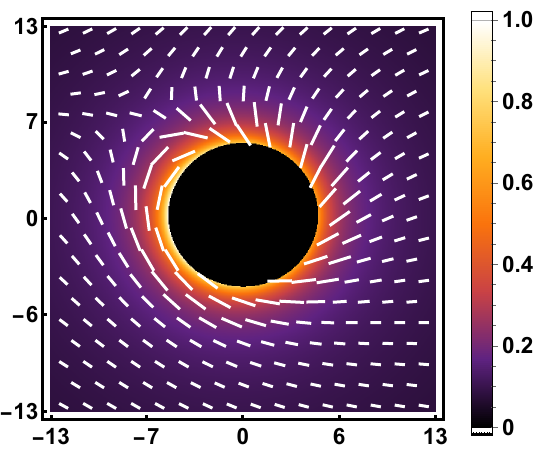}}
	\subfigure[$\theta_{\mathrm{o}}=55^\circ,n_{\mathrm{p}}=1.2$]{\includegraphics[scale=0.4]{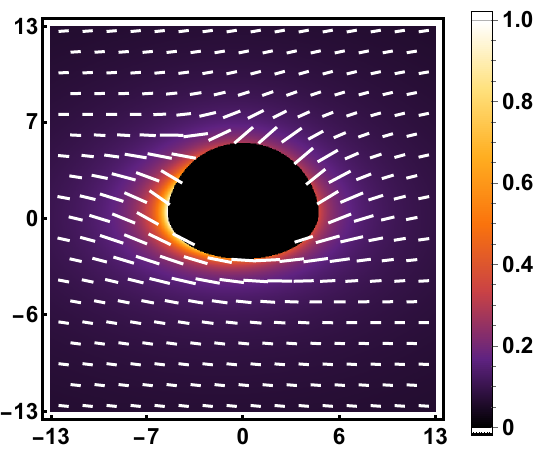}}
	\subfigure[$\theta_{\mathrm{o}}=80^\circ,n_{\mathrm{p}}=1.2$]{\includegraphics[scale=0.4]{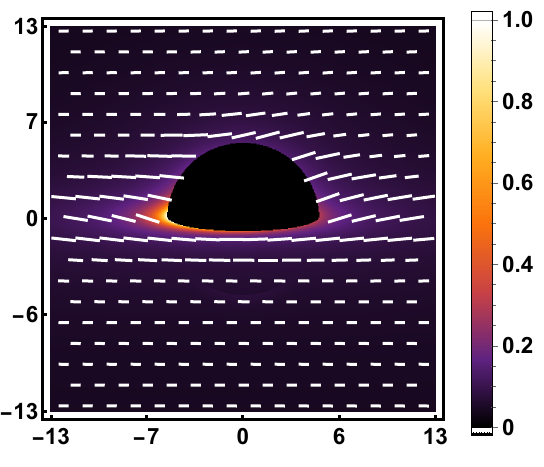}}
	
	\subfigure[$\theta_{\mathrm{o}}=0^\circ,n_{\mathrm{p}}=1.3$]{\includegraphics[scale=0.4]{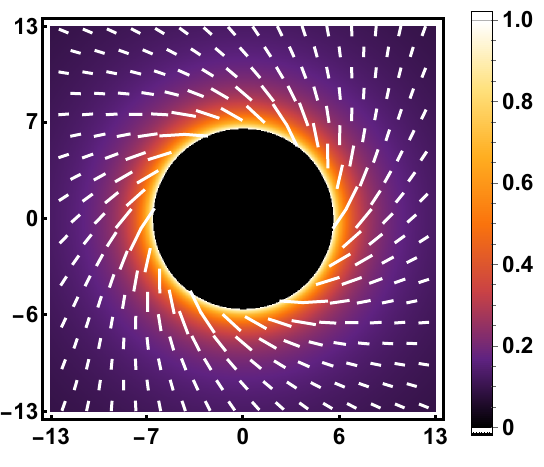}}
	\subfigure[$\theta_{\mathrm{o}}=20^\circ,n_{\mathrm{p}}=1.3$]{\includegraphics[scale=0.4]{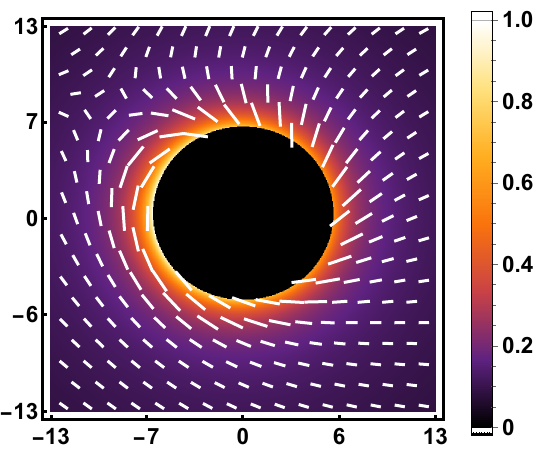}}
	\subfigure[$\theta_{\mathrm{o}}=55^\circ,n_{\mathrm{p}}=1.3$]{\includegraphics[scale=0.4]{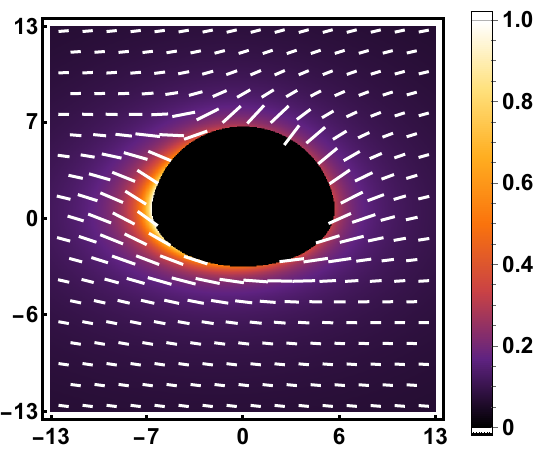}}
	\subfigure[$\theta_{\mathrm{o}}=80^\circ,n_{\mathrm{p}}=1.3$]{\includegraphics[scale=0.4]{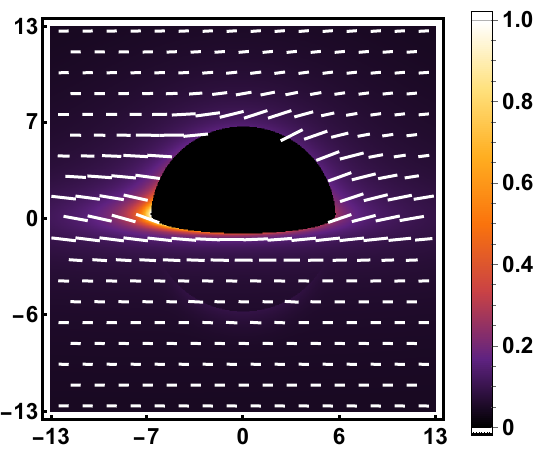}}
	
	\subfigure[$\theta_{\mathrm{o}}=0^\circ,n_{\mathrm{p}}=1.4$]{\includegraphics[scale=0.4]{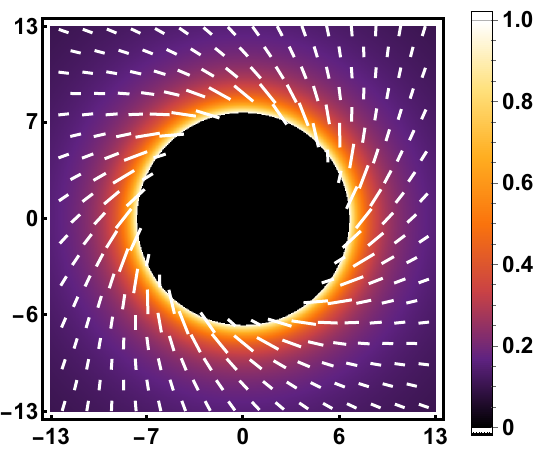}}
	\subfigure[$\theta_{\mathrm{o}}=20^\circ,n_{\mathrm{p}}=1.4$]{\includegraphics[scale=0.4]{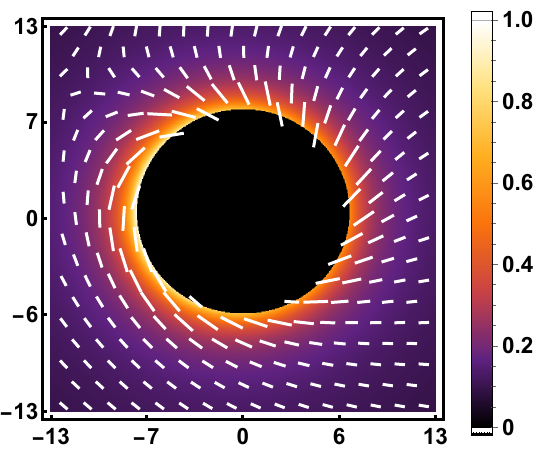}}
	\subfigure[$\theta_{\mathrm{o}}=55^\circ,n_{\mathrm{p}}=1.4$]{\includegraphics[scale=0.4]{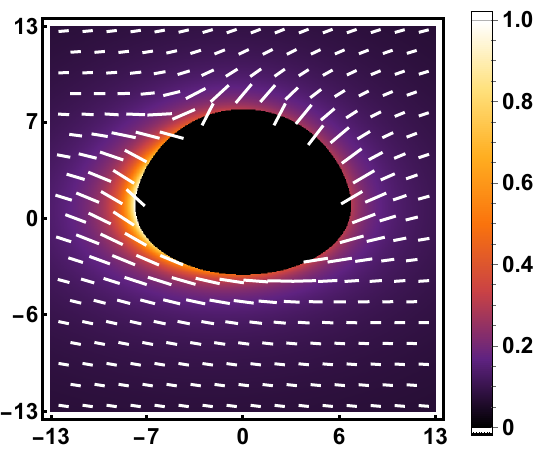}}
	\subfigure[$\theta_{\mathrm{o}}=80^\circ,n_{\mathrm{p}}=1.4$]{\includegraphics[scale=0.4]{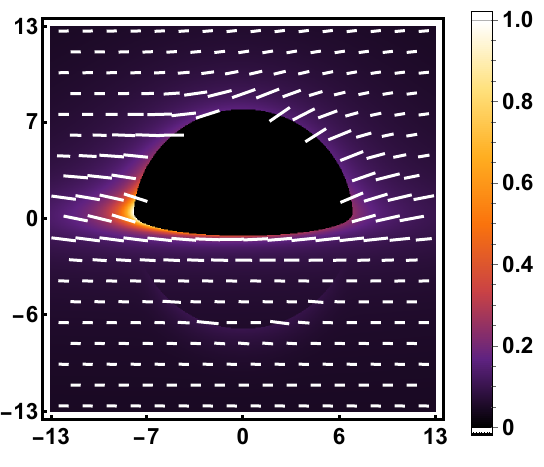}}
	
	\caption{Polarization images of the neutron star. The short white segments represent the linear polarization vectors. Their lengths and directions encode the linearly polarized intensity $P_{\mathrm{obs}}$ and the EVPA $\chi_{\mathrm{evpa}}$, respectively. The fixed parameters are $\vec{B}=(0.87,0.5,0)$ and $r_{\mathrm{o}}=200$.}
	\label{fig1}
\end{figure}

Fig.~\ref{fig1} presents the numerical results for observer inclinations $\theta_{\mathrm{o}}=0^\circ,20^\circ,55^\circ,80^\circ$ and polytropic indices $n_{\mathrm{p}}=1.1,1.2,1.3,1.4$. The background shows the intensity distribution of a neutron star illuminated by a thin accretion disk. Because the neutron star surface is assumed to absorb incident photons completely, the black region corresponds to the stellar silhouette. The intensity reaches its maximum near the neutron star surface and decreases rapidly away from the star. The short white segments represent the linear polarization vectors $\vec{f}$. Their lengths and directions indicate the linearly polarized intensity $P_{\mathrm{obs}}$ and the EVPA $\chi_{\mathrm{evpa}}$, respectively. The polarized intensity is generally positively correlated with the total intensity. It is substantially stronger in bright regions than in dark regions, and $P_{\mathrm{obs}}$ reaches its maximum near the neutron star surface. With the adopted absorption assumption, polarization signals from the stellar interior are excluded from the calculation, so no linear polarization vectors $\vec{f}$ are shown in that region.

\begin{figure}[!htbp]
	\centering 
	\subfigure[Horizontal direction]{\includegraphics[scale=0.7]{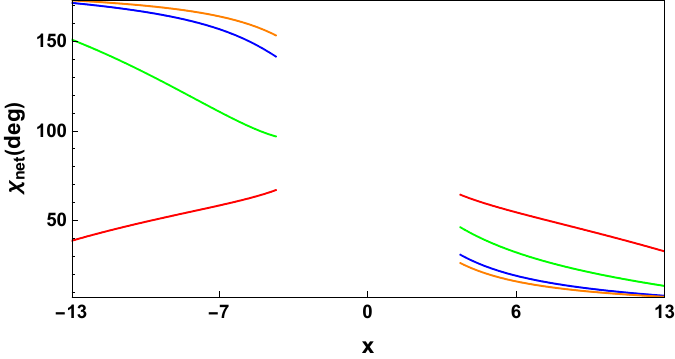}}
	\subfigure[Vertical direction]{\includegraphics[scale=0.7]{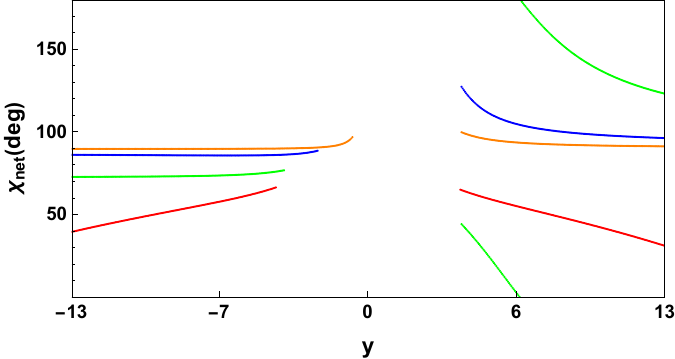}}
	
	\caption{Horizontal and vertical profiles of $\chi_{\mathrm{net}}$ at different observer inclinations. The fixed parameters are $\vec{B}=(0.87,0.5,0)$ and $n_{\mathrm{p}}=1.1$. The red, green, blue, and orange curves correspond to $\theta_{\mathrm{o}}=0^\circ,20^\circ,55^\circ,80^\circ$, respectively.}
	\label{fig2}
\end{figure}

Fig.~\ref{fig2} quantitatively shows the horizontal and vertical profiles of $\chi_{\mathrm{net}}$ at different observer inclinations for $n_{\mathrm{p}}=1.1$. Each curve has two branches because there is no polarization signal inside the neutron star silhouette. When $\theta_{\mathrm{o}}=0^\circ$, the $\chi_{\mathrm{net}}$ profiles are approximately symmetric about the image center in both the horizontal and vertical directions and correspond to a counterclockwise pattern. As $\theta_{\mathrm{o}}$ increases, the left-right symmetry of the horizontal profile is progressively broken. The vertical polarization pattern changes from counterclockwise to nearly azimuthal or clockwise. Notably, the curve for $\theta_{\mathrm{o}}=20^\circ$ exhibits a jump from $0^\circ$ to $180^\circ$ near $y\simeq6$. This jump arises because $\chi_{\mathrm{net}}$ is defined modulo $180^\circ$ and does not represent a physical discontinuity in the polarization direction.

\begin{figure}[!htbp]
	\centering 
	\subfigure{\includegraphics[scale=0.7]{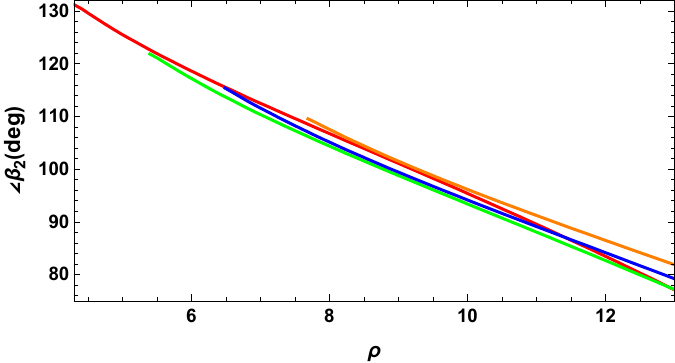}}
	
	\caption{$\angle\beta_2$ as a function of the polar radius $\rho$. The fixed parameters are $\vec{B}=(0.87,0.5,0)$ and $\theta_{\mathrm{o}}=0^\circ$. The red, green, blue, and orange curves correspond to $n_{\mathrm{p}}=1.1,1.2,1.3,1.4$, respectively.}
	\label{fig3}
\end{figure}

For polarization images with approximate central symmetry, the overall polarization structure can be characterized by $\angle\beta_2$. Fig.~\ref{fig3} shows how $\angle\beta_2$ varies with the polar radius $\rho$ for different polytropic indices $n_{\mathrm{p}}$ at fixed $\theta_{\mathrm{o}}=0^\circ$. All curves decrease monotonically with $\rho$ and remain between $0^\circ$ and $180^\circ$. The mean polarization direction within each annulus therefore shifts gradually from an approximately azimuthal pattern toward the radial direction. The mean deflection remains counterclockwise throughout, and the polarization pattern does not reverse. Overall, $\angle\beta_2$ exhibits a stable and continuous variation with radius.

\begin{figure}[!htbp]
	\centering 
	\subfigure[$\vec{B}=(0.97,0.26,0)$]{\includegraphics[scale=0.4]{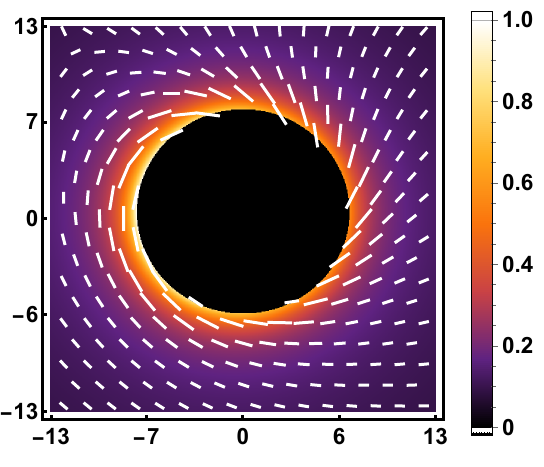}}
	\subfigure[$\vec{B}=(1,0,0)$]{\includegraphics[scale=0.4]{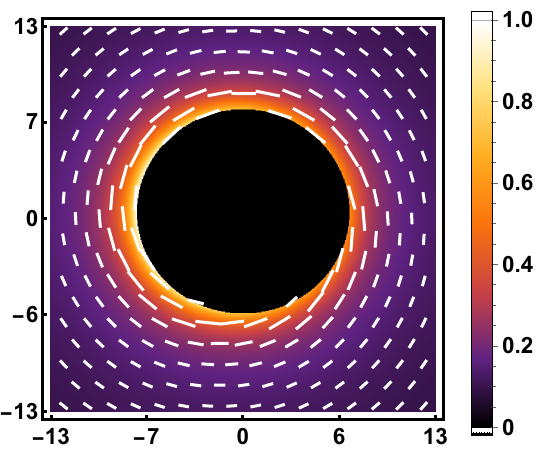}}
	\subfigure[$\vec{B}=(0,1,0)$]{\includegraphics[scale=0.4]{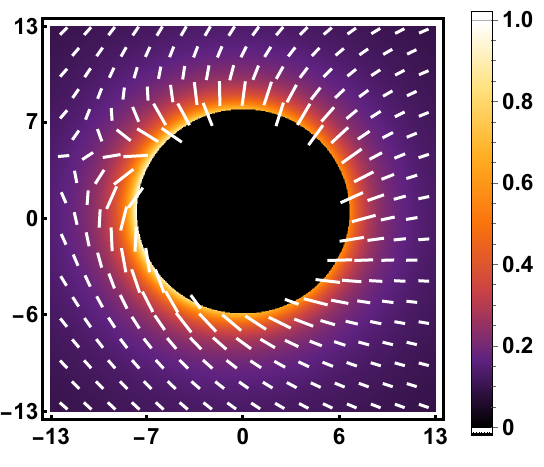}}
	\subfigure[$\vec{B}=(0,0,1)$]{\includegraphics[scale=0.4]{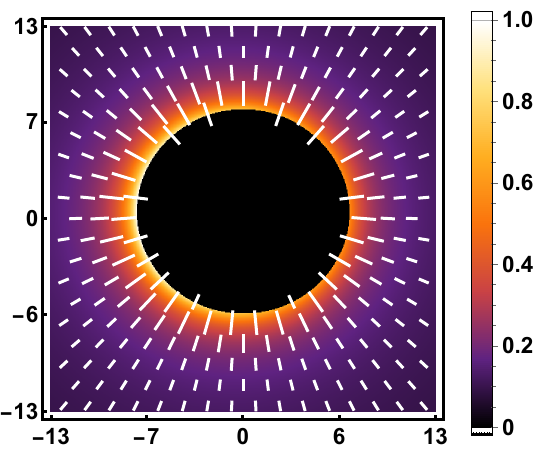}}
	
	\subfigure[$\vec{B}=(0.97,0.26,0)$]{\includegraphics[scale=0.4]{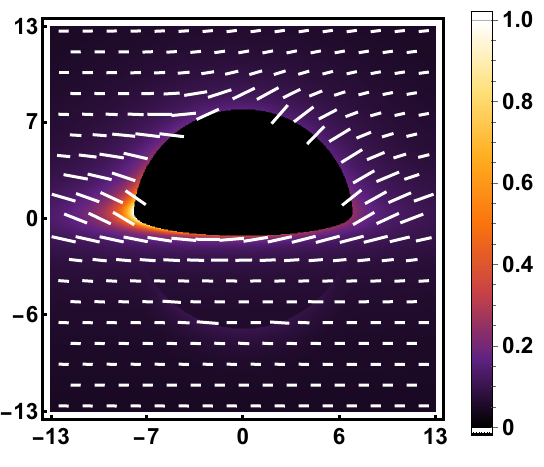}}
	\subfigure[$\vec{B}=(1,0,0)$]{\includegraphics[scale=0.4]{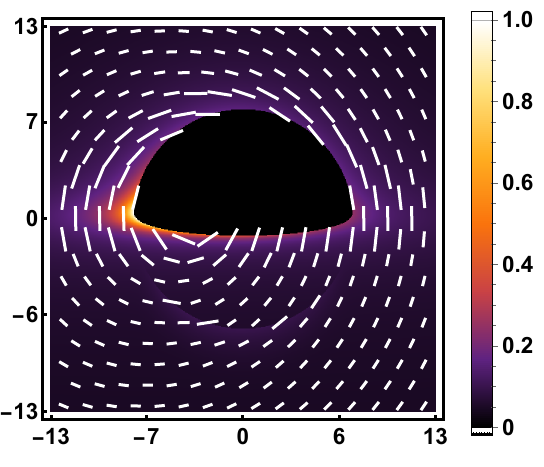}}
	\subfigure[$\vec{B}=(0,1,0)$]{\includegraphics[scale=0.4]{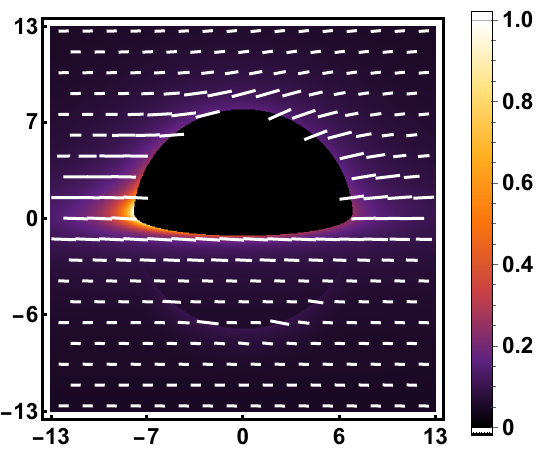}}
	\subfigure[$\vec{B}=(0,0,1)$]{\includegraphics[scale=0.4]{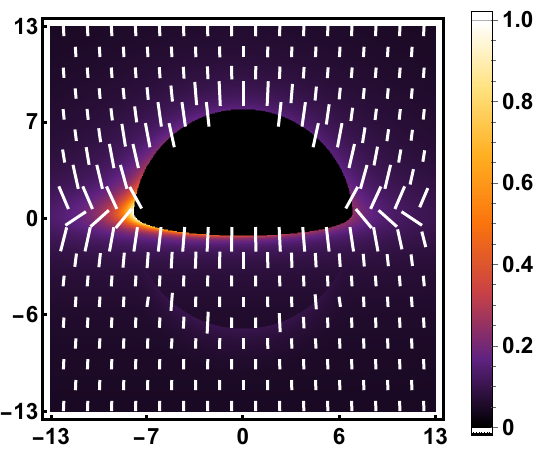}}

	\caption{Effects of different magnetic field configurations on neutron star polarization images. The first and second rows correspond to $\theta_{\mathrm{o}}=20^\circ$ and $\theta_{\mathrm{o}}=80^\circ$, respectively. The fixed parameters are $n_{\mathrm{p}}=1.4$ and $r_{\mathrm{o}}=200$.}
	\label{fig4}
\end{figure}

The above analysis is based on the magnetic field configuration $\vec{B}=(0.87,0.5,0)$ in the equatorial plane. Fig.~\ref{fig4} presents polarization images for several other configurations, all with the polytropic index fixed at $n_{\mathrm{p}}=1.4$. The image for $\vec{B}=(0.97,0.26,0)$ is similar to that for $\vec{B}=(0.87,0.5,0)$. For the lower observer inclination $\theta_{\mathrm{o}}=20^\circ$ shown in the first row, the polarization vectors are approximately azimuthal for $\vec{B}=(1,0,0)$. In contrast, they are clearly radial for $\vec{B}=(0,0,1)$. Different magnetic field configurations therefore have a pronounced effect on the polarization images. This effect is mainly reflected in the polarization direction, while the overall polarized intensity distribution changes comparatively little. The polarization direction can thus serve as an important probe of the magnetic field configuration.

\begin{figure}[!htbp]
	\centering 
	\subfigure[BH, $\theta_{\mathrm{o}}=20^\circ$]{\includegraphics[scale=0.4]{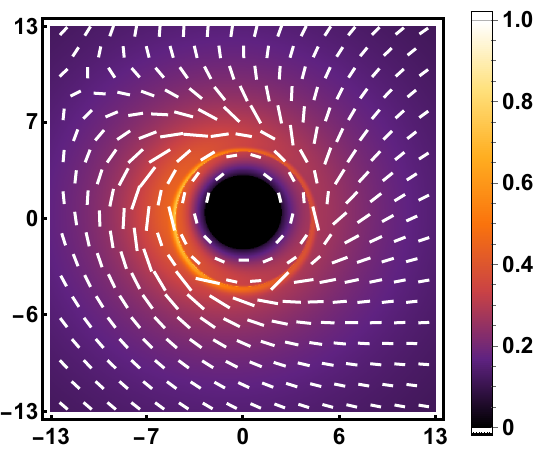}}
	\subfigure[NS, $\theta_{\mathrm{o}}=20^\circ$]{\includegraphics[scale=0.4]{p14.pdf}}
	\subfigure[BH, $\theta_{\mathrm{o}}=80^\circ$]{\includegraphics[scale=0.4]{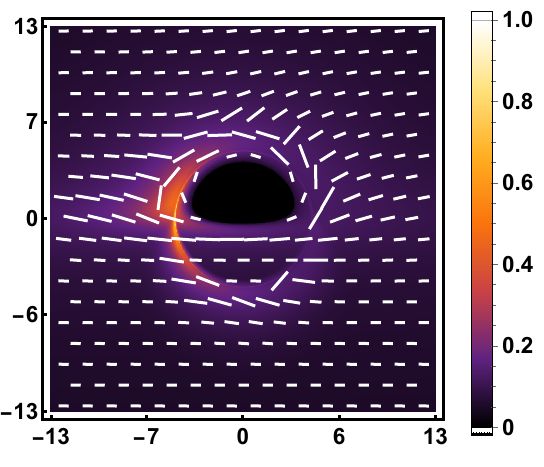}}
	\subfigure[NS, $\theta_{\mathrm{o}}=80^\circ$]{\includegraphics[scale=0.4]{p16.pdf}}
	
	\caption{Comparison between polarization images of a Schwarzschild black hole (BH) and a neutron star (NS). The neutron star polytropic index is fixed at $n_{\mathrm{p}}=1.4$, the magnetic field configuration is $\vec{B}=(0.87,0.5,0)$, and all other plotting parameters are identical.}
	\label{fig5}
\end{figure}

Finally, we compare the polarization images of a Schwarzschild black hole and a neutron star. As shown in Fig.~\ref{fig5}, the polarized intensity of each object is closely related to its total intensity distribution, and brighter regions generally exhibit stronger linearly polarized signals. For the Schwarzschild black hole, no polarization signal appears in the central dark region due to the presence of the event horizon~\cite{yang2026observational}. Under our absorption assumption, integration terminates when a ray reaches the neutron star surface, so the stellar silhouette likewise contains no polarization signal. Although the exterior of the neutron star is described by the Schwarzschild metric, the polarization images of the two objects remain clearly different. At low observer inclination, the polarized signal in the black hole image is concentrated near the photon ring, whereas the polarized intensity of the neutron star peaks near the stellar surface. At high observer inclination, a clear lensed image appears in the lower part of the black hole image. The lengths and directions of its polarization vectors differ markedly from those in the corresponding region of the neutron star image. For the same mass, the central dark region of the neutron star is also substantially larger than that of the black hole.

\section{Conclusions and Discussion}\label{sec5}

We have investigated linear polarization images of static, spherically symmetric neutron stars illuminated by a thin accretion disk. We constructed the equilibrium neutron star configurations with a polytropic EOS and combined ray tracing with parallel transport of the linear polarization vector to calculate the observed intensity $I_{\mathrm{obs}}$, total polarized intensity $P_{\mathrm{obs}}$, and EVPA $\chi_{\mathrm{evpa}}$. In our model, the neutron star interior is optically opaque to electromagnetic radiation. We therefore terminate the integration when a ray reaches the neutron star surface at $r=R_{\star}$ and consider radiative propagation only in the exterior spacetime. We also introduced the ``net EVPA'' $\chi_{\mathrm{net}}$ and the second azimuthal Fourier mode $\angle\beta_2$ to quantify the direction of linear polarization on the image plane and its overall radial evolution.

The numerical results show that the total polarized intensity is generally positively correlated with the total intensity distribution. Brighter regions usually exhibit stronger linearly polarized signals, and $P_{\mathrm{obs}}$ reaches its maximum near the neutron star surface. The stellar silhouette contains no polarization signal because of the imposed boundary condition. The observer inclination strongly affects the image symmetry and polarization direction. At $\theta_{\mathrm{o}}=0^\circ$, the polarization image is approximately centrally symmetric. As $\theta_{\mathrm{o}}$ increases, the left-right symmetry of the horizontal profile is progressively broken, while the vertical polarization pattern changes from counterclockwise to nearly azimuthal or clockwise. We further analyzed the approximately centrally symmetric images at $\theta_{\mathrm{o}}=0^\circ$. For every polytropic index $n_{\mathrm{p}}$ considered, $\angle\beta_2$ decreases monotonically with the polar radius $\rho$ while maintaining a mean counterclockwise rotation. We also examined the effect of the magnetic field configuration $\vec{B}=\vec{B}(B_r,B_\phi,B_z)$. The field $\vec{B}$ primarily changes the direction of the polarization vectors and has a comparatively limited effect on the overall polarized intensity distribution. At low observer inclination, the purely radial field $\vec{B}=(1,0,0)$ produces an approximately azimuthal polarization pattern, whereas the field $\vec{B}=(0,0,1)$ perpendicular to the equatorial plane produces a clearly radial pattern.

The comparison between the Schwarzschild black hole and neutron star shows that the two compact objects have distinct polarization signatures even though the neutron star exterior is described by the Schwarzschild metric. At low inclination, the strongly polarized region of the black hole is concentrated near the photon ring, whereas that of the neutron star lies near the stellar surface. At high inclination, a clear lensed image appears below the black hole image, and the lengths and directions of its polarization vectors differ markedly from those in the corresponding region of the neutron star image. The location of the strongly polarized region, the polarization structure of the lensed image at high inclination, and the size of the central dark region without a polarization signal can therefore serve as important observational signatures for distinguishing black holes from neutron stars.

Overall, our results provide theoretical support for distinguishing neutron stars from black holes through polarization images. Rather than relying on intensity images alone, more reliable diagnostics can be obtained by combining intensity information with polarization signatures. Future work will consider neutron stars described by different EOSs and incorporate more realistic polarized radiative transfer models. These extensions will help further clarify the polarization signatures of different compact objects and the physical origins of their differences.


\cleardoublepage

\vspace{10pt}
\noindent {\bf Acknowledgments}

\noindent
This work is supported by the National Natural Science Foundation of China (Grants Nos. 12675078).

\bibliographystyle{utphys} 
\bibliography{biblio} 

\end{document}